\documentclass[letterpaper,showpacs,preprintnumbers,amsmath,amssymb,nofootinbib,twocolumn,superscriptaddress,notitlepage,prl]{revtex4-1}

\usepackage{cancel}

\usepackage{accents}
\usepackage{mciteplus,slashed}
\usepackage{amssymb,cancel,amsmath}
\usepackage{dcolumn}% Align table columns on decimal point
\usepackage{bm}% bold math
\usepackage[caption=false]{subfig}
\usepackage{appendix}
\usepackage{feynmp-auto}
\usepackage[T1]{fontenc}	
\usepackage{csvsimple}
\usepackage[colorlinks=true,citecolor=BrickRed,linkcolor=blue,urlcolor=RoyalBlue]{hyperref}
\usepackage[section]{placeins}
\usepackage[capitalise]{cleveref}
\usepackage{booktabs}
\usepackage{graphicx}
\usepackage{mathrsfs}
\usepackage{multirow}
\usepackage[utf8]{inputenc}
\usepackage{siunitx}
\usepackage{comment}
\usepackage{comment}

\usepackage[dvipsnames]{xcolor}
\usepackage[normalem]{ulem}

\begin{document}

\title{PeV Tau Neutrinos to Unveil Ultra-High-Energy Sources}

\author{Carlos~A.~Arg{\"u}elles}
\email{carguelles@fas.harvard.edu}
\affiliation{Department of Physics \& Laboratory for Particle Physics and Cosmology, Harvard University, Cambridge, MA 02138, USA}

\author{Francis~Halzen}
\email{halzen@icecube.wisc.edu}
\affiliation{Department of Physics \& Wisconsin IceCube Particle Astrophysics Center, University of Wisconsin, Madison, WI 53706, USA}

\author{Ali~Kheirandish}
\email{kheirandish@psu.edu}
\affiliation{Department of Physics, Department of Astronomy \& Astrophysics, \& Center for Multimessenger Astrophysics, Institute for Gravitation and the Cosmos, The Pennsylvania State University, University Park, PA 16802, USA}

\author{Ibrahim~Safa}
\email{isafa@fas.harvard.edu}
\affiliation{Department of Physics \& Laboratory for Particle Physics and Cosmology, Harvard University, Cambridge, MA 02138, USA}
\affiliation{Department of Physics \& Wisconsin IceCube Particle Astrophysics Center, University of Wisconsin, Madison, WI 53706, USA}

\date{\today}

\begin{abstract}
The observation of ultra-high-energy EeV-energy cosmogenic neutrinos provides a direct path to identifying the sources of the highest energy cosmic rays; searches have so far resulted in only upper limits on their flux. However, with the realization of cubic-kilometer detectors such as IceCube and, in the near future, KM3NeT, GVD-Baikal, and similar instruments, we anticipate the observation of PeV-energy cosmic neutrinos with high statistics. In this context, we draw attention to the opportunity to identify EeV tau neutrinos at PeV energy using Earth-traversing tau neutrinos. We show that Cherenkov detectors can improve their sensitivity to transient point sources by more than an order of magnitude by indirectly observing EeV tau neutrinos with initial energies that are nominally beyond their reach. This new technique also improves their sensitivity to the ultra-high-energy diffuse neutrino flux by up to a factor of two. Our work exemplifies how observing tau neutrinos at PeV energies provides an unprecedented reach to EeV fluxes.
\end{abstract}

\maketitle

\textbf{\textit{Introduction.\label{sec:introduction}---}} Ultra-high-energy (UHE) cosmic-rays have been detected with energies approaching $10^{21}~{\rm eV}$. This endpoint could reflect either the ultimate reach in energy of the cosmic accelerators, or it could be a signature of a cosmic-ray–opaque Universe, where the UHE cosmic-ray flux is absorbed by interactions with microwave photons. This is the origin of  the so-called cosmogenic---or Greisen–Zatsepin–Kuzmin (GZK)~\cite{Greisen:1966jv, Zatsepin:1966jv, Beresinsky:1969qj}---neutrino flux from the decay of pions produced in the dominant process $p\gamma \to \Delta^+ \to n\pi^+/p\pi^0$.

GZK neutrinos have been the target of km-scale~\cite{Roberts:1992re, Halzen:2008zz} optical neutrino detectors, such as the IceCube Neutrino Observatory at the South Pole~\cite{Aartsen:2016nxy}; very large arrays of water tanks, such as the Pierre Auger observatory in the Argentinian Pampas~\cite{ThePierreAuger:2015rma}; and antennas that detect the radio emission from neutrino-initiated showers, suspended from high-altitude balloons ~\cite{Gorham:2008dv} or deployed either on ~\cite{Barwick:2014rca} or below the Antarctic ice sheet~\cite{Allison:2011wk}, among others. This multipronged approach to discover GZK neutrinos has yielded only upper limits on their flux so far ~\cite{Abbasi:2010ak, Barwick:2014pca, Aab:2015kma, Aartsen:2016ngq, Allison:2015eky,Allison:2018cxu,Aartsen:2018vtx,Aab:2019auo}.

The observation of this so-called ``guaranteed'' flux is one of the primary targets of next-generation neutrino observatories---such as IceCube-Gen2, RNO-G, TAMBO, POEMMA, and others~\cite{Aguilar:2019jay, Fang:2017mhl, Neronov:2016zou, Krizmanic:2019hiq, Aartsen:2014njl}---because it will shed light on the composition of UHE cosmic rays \cite{Ahlers:2010fw,Ahlers:2012rz}, their sources and cosmological evolution \cite{Berezinsky:2002nc,Fodor:2003ph,Yuksel:2006qb,Takami:2007pp}, and the properties of the extragalactic background light \cite{Hooper:2004jc,Ave:2004uj,Hooper:2006tn,Allard:2006mv,Anchordoqui:2007fi,Kotera:2010yn,Decerprit:2011qe,Aloisio:2009sj, Ahlers:2011sd,Ahlers:2012rz}.
The nonobservation of GZK neutrinos has already provided valuable information along these lines.
These experiments use a common strategy to search for GZK neutrinos, which primarily relies upon the observation of very high energy events near the horizon, often referred to as Earth-skimming neutrinos~\cite{Kotera:2010yn, Jeong:2017mzv, Dutta:2002zc, Venters:2019xwi, Reno:2019jtr, Dutta:2000jv, Dutta:2005yt}.
This is because Earth is opaque to neutrinos at EeV energies; the survival probability of a primary neutrino of 100\,PeV energy traversing Earth is 0.2 for $\cos\theta = -0.1$ (Earth-skimming), $2 \times 10^{-6}$ for $\cos\theta = -0.6$ (Earth-mantle-crossing), and $10^{-19}$ for $\cos\theta = -1$ (Earth-core-crossing), where $\theta$ is the zenith angle.

However, neutrinos that experience charged-current interactions produce high-energy charged leptons. These leptons will undergo catastrophic energy losses in Earth until they reach energies such that their interaction and decay lengths are comparable, at which point they produce secondary neutrinos. As pointed out for the first time in~\cite{Ritz:1987mh} and recently revisited in~\cite{Safa:2019ege}, this process is particularly important for tau neutrinos.
Taus have a short lifetime and decay promptly into a secondary tau neutrino that carries a large fraction of the primary energy. Moreover, it was pointed out in \cite{Beacom:2001xn} that the decay of the tau also contributes a non-negligible flux of secondary electron and muon antineutrinos. This process is commonly referred to as \textit{tau regeneration}.

In this {\em letter}, we take advantage of the tau regeneration process to identify a complementary way to search for UHE neutrinos and discuss the connection between current PeV and forthcoming EeV neutrino measurements considering both the observation of transient sources and their diffuse flux. We find that the current point-source sensitivity to UHE neutrinos relying on muon event selections is enhanced by more than an order of magnitude for Earth-traversing directions when tau regeneration is taken into account. For the observation of the diffuse flux, we illustrate the role of tau regeneration assuming alternatively an astrophysical power-law flux and a conventional GZK spectrum.
We find that the two components can be separated by measurement of their angular and energy distributions.
Using this information, the sensitivity to specific GZK neutrino models is improved by up to a factor of two compared to current techniques.
Therefore, our approach impacts both the characterization of the diffuse spectra and the discovery potential of neutrino sources for detectors that operate in the PeV energy range.

\textbf{\textit{Appearance of EeV neutrinos at PeV energy.\label{sec:theory}---}}
Though neutrinos rarely interact, their interaction cross section with nucleons grows linearly with energy up to approximately 1~TeV, after which the typical momentum transferred exceeds the weak vector-boson masses.
Above this energy, the cross-section growth slows down, ultimately growing only as $\log^2 s$~\cite{Block:2014kza}, where $s$ is the center-of-mass energy squared.
This scaling is a universal characteristic of the deep-inelastic proton-, neutrino-, and photon-nucleon scattering~\cite{Arguelles:2015wba}.
The distance that a neutrino traverses Earth to reach an underground detector eventually exceeds or is comparable to the corresponding neutrino interaction length, e.g. $\sim2,200$~km at 1~PeV and $\sim100$~km at 1~EeV.
The antineutrino interaction lengths are comparable because the interaction is dominated by sea quarks and anti quarks in the proton~\cite{Gandhi:1995tf}.
The ratio of neutrino charged-to-neutral current interactions is approximately 2.5~\cite{Gandhi:1995tf} in this energy range, which implies that more often than not, high-energy charged leptons will be produced in neutrino interactions. Alternatively, in the case of neutral-current interactions, the neutrino loses, on average, 30\% of its initial energy.

The secondary charged leptons lose energy catastrophically.
At EeV energies, taus lose energy predominantly through photohadronic processes and pair production~\cite{Chirkin:2004hz}. 
For electrons, bremsstrahlung dominates, while for muons the dominant energy loss processes are pair production and bremsstrahlung~\cite{Koehne:2013gpa} with a combined contribution that approximately matches the photohadronic processes~\cite{Bugaev:2002gy}.

While muons will lose most of their energy before decaying, taus will decay without significant energy loss below 100 PeV, resulting in neutrinos with energies close to the parent neutrino energy.
An EeV tau neutrino will be subject to, on average, three neutral-current and five charged-current interactions prior to crossing Earth's diameter and reaching the detector; its typical final energy is between 10~TeV and 10~PeV.
\begin{figure}[ht!]
    \centering
    \includegraphics[width=\columnwidth]{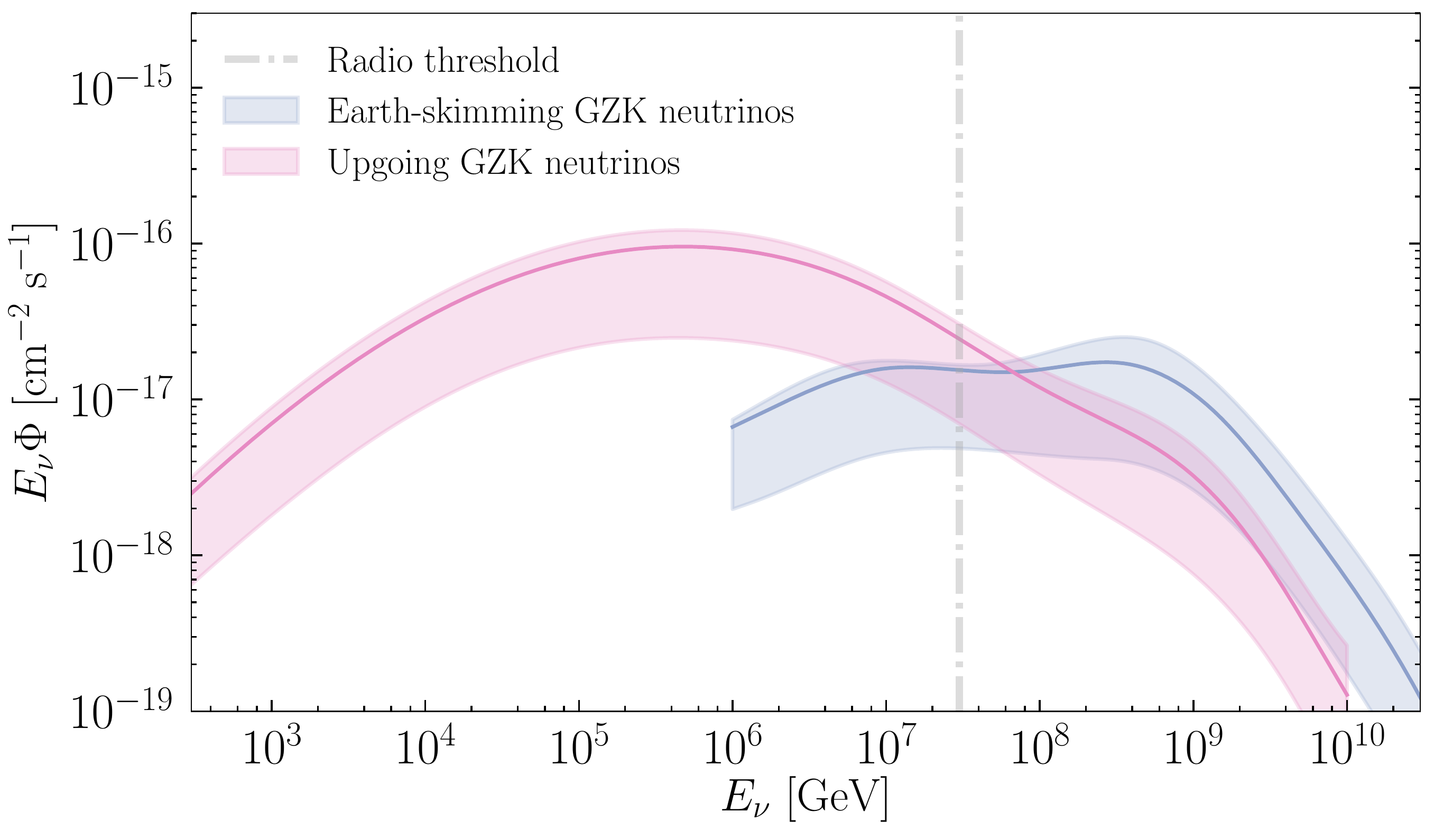}
    \caption{\textbf{\textit{At-detector angle-integrated neutrino flux intensity.}}
    EeV neutrinos pile up at PeV energies after propagating through Earth (pink).
    We integrate over the respective solid angles and define Earth-skimming as within 10$^{\circ}$ of the horizon.
    Upgoing neutrinos are accessible to water- and ice-Cherenkov detectors but fall below the threshold for radio experiments (shown as a vertical dot-dashed gray line). As an example of a baseline GZK model (blue), we use the flux given by Ahlers \textit{et. al.} 2010 \cite{Ahlers:2010fw}.}
    \label{fig:fluxes}
\end{figure}
Fig.~\ref{fig:fluxes} shows the GZK neutrino fluxes before (blue) and after (pink) their propagation through Earth.
In this figure, we have scaled the flux by the neutrino energy such that the lines are approximately proportional to the event rate.
As anticipated, before Earth propagation, the flux is maximal between $10^8$ and $10^9$~GeV, while after propagation it peaks between $10^5$ to $10^6$~GeV. 

\textbf{\textit{Observing sources of UHE neutrinos.\label{sec:source_analysis}---}} Astrophysical beam dumps in the vicinity of the sources of UHE cosmic rays provide the opportunity for the production of charged and neutral pions, and therefore, of neutrinos with energies exceeding tens of PeV. Potential sources of UHE neutrinos include binary neutron star (BNS) mergers ~\cite{Kimura:2017kan}, magnetars \cite{Fang:2018hjp,Murase:2009pg, Carpio:2020wzg}, young and fast spinning pulsars~\cite{Fang:2013vla}, supermassive black hole mergers~\cite{Yuan:2020oqg}, gamma-ray burst (GRB) afterglows \cite{Murase:2007yt}, and blazars with a luminous dust-torus ~\cite{Murase:2014foa,Petropoulou:2016ujj,Keivani:2018rnh}. The predicted neutrino energy spectrum peaks in the range of 10--100~PeV, energies that are typically below the sensitivity of the current generation of neutrino telescopes.
However, the presence of a flux extending beyond 10~PeV presents an opportunity to probe the sources via the regenerated flux from tau neutrinos.

\begin{figure}[t!]
    \centering
    \includegraphics[width=\columnwidth]{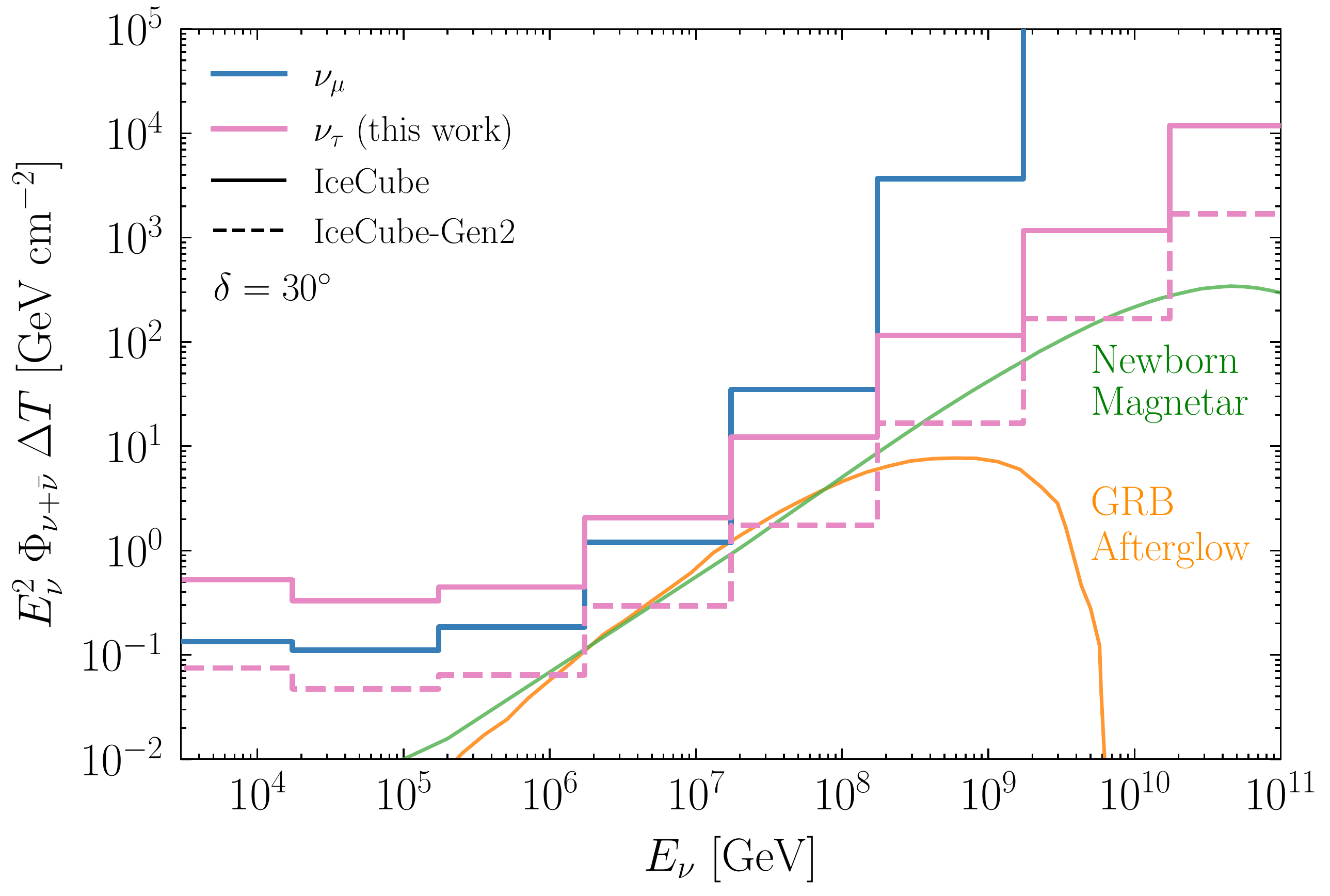}
    \caption{\textbf{\textit{Improved point source sensitivity with $\nu_{\tau}$.}}
    Differential sensitivity to an $E^{-2}$ time-integrated flux for transient UHE sources at a fixed declination ($\delta = 30^\circ$). The current sensitivity of IceCube is shown in blue, while the sensitivity to muons from tau neutrinos is shown in pastel pink. The projected sensitivity resulting from our method for IceCube-Gen2 (optical-only) is shown as a dashed line.
    To illustrate the power of our method, we compare it to the predicted time-integrated neutrino flux from %binary neutron star mergers \cite{Fang:2017mhl}
    newborn magnetars (green) \cite{Carpio:2020wzg} for a distance of 0.3 Mpc and GRB afterglows (orange) \cite{Murase:2007yt} at 20 Mpc. 
    }
    \label{fig:transients}
\end{figure}

In order to demonstrate the power of the regeneration-based technique for identifying sources of UHE neutrinos, we compute the differential sensitivity of IceCube for a neutrino flux in the range of 1 PeV to 1 ZeV.
Neutrinos with this energy traversing Earth will generate a cascade of muon and tau neutrino fluxes at the detector.
In contrast to the primary all-flavor flux that is attenuated, the regenerated flux results in a substantial intensity of neutrinos of PeV energy, where the sensitivity is optimal for detection by optical Cherenkov detectors.
Here, we utilize the IceCube differential sensitivity for transients~\cite{IceCube:2020mzw}.
For each energy decade, we inject a flux of tau neutrinos, assuming the spectrum follows a $E^{-2}$ distribution.
To obtain the normalization for each bin, we determine the corresponding flux of secondary neutrinos that reaches the detector.
The sensitivity of IceCube to a flux of muons arriving at the horizon constrains the flux that arrives at the detector in any direction, when excluding Earth effects.
Thus, we use the sensitivity at $\delta = 0$ to constrain the initial flux at higher energies; that is, we constrain the normalization of the injected spectrum at different declinations by folding the sensitivity at the horizon with the expected number of muons from tau decay produced by the propagated secondary neutrinos.
In Fig.~\ref{fig:transients}, we show the corresponding differential sensitivity of IceCube to the time-integrated flux in the energy range of $10^4$~GeV -- $10^{11}$~GeV.
For instance, for a declination of 30$^{\circ}$, the $\nu_\tau$ sensitivity to a transient outperforms the conventional time-dependent search with $\nu_\mu$ for energies greater than $10^7$~GeV.
The improvement grows with energy, exceeding an order of magnitude at $10^9$~GeV.
We also show the corresponding sensitivity for IceCube-Gen2 demonstrating the enhanced reach of time-dependent searches with water-Cherenkov telescopes to sources of UHE neutrinos.
The significance of this improvement is underscored by the power of this methodology to probe neutrinos from GRB afterglows and newborn magnetars, which is out of reach for conventional time-dependent follow-up.
We note that the relative improvement in the sensitivity increases with declination, and reaches more than three orders of magnitude at $10^{9}$ GeV and 60 degrees below the horizon. The method creates new opportunities for the discovery of cosmic accelerators that produce the most energetic particles and were previously inaccessible to Cherenkov detectors.

\textbf{\textit{Diffuse Fluxes.\label{sec:analysis}---}}
We illustrate the importance of tau regeneration in searching for GZK neutrinos with an illustrative analysis. 
We evaluate the response of IceCube to neutrinos with energy above 100~TeV, where the atmospheric neutrino component is subdominant~\cite{Abbasi:2020jmh}. The muon-neutrino effective area is taken from~\cite{Aartsen:2016xlq}. We consider two isotropic components: an astrophysical component modeled by a power-law whose parameters have been determined using an independent cascade set of data and a GZK spectrum that we model according to~\cite{Ahlers:2010fw}.
The two components can be differentiated by their energy dependence as well as their angular distribution.
This difference is caused by a contrasting amount of absorption as a function of energy and traversed column depth.

\begin{comment}
\begin{table}%[!ht]
    \label{tab:Jtable}
    \makebox[\linewidth]{
    \begin{tabular}{ r c l }
    \hline
    & {\bf Model Rejection} & {\bf Factors} \\
    \hline \hline
        \textbf{Time}  & \textbf{Horizon} & \textbf{Horizon+Upgoing}  \\
         years & $\delta \leq 8^\circ$ &  $\delta \leq 90^\circ$ \\ \hline
        1  & $1.65$ & $0.77$ \\\hline 
        5 & $0.64$ & $0.27$  \\ \hline
        10 & $0.42$ & $0.17$    \\ \hline \hline
        
    \end{tabular}
    }
    % %\internallinenumbers
    \caption{
    \textbf{\textit{Sensitivity improvement for GZK neutrino searches.}}
    Each row corresponds to the number of years of data taking for IceCube-Gen2.
    The middle column shows the model rejection factor when measuring neutrinos near the horizon, while the rightmost column lists the corresponding number when measuring both horizon and upgoing events.
    The model rejection factor is defined as the ratio of the baseline GZK flux normalization with respect to the limit obtained in this analysis at 90\%~CL. 
    As a baseline GZK fit we use the flux given by Ahlers \textit{et. al.} 2010 \cite{}.
    }
\end{table}
\end{comment}
%
To quantify the significance of the Earth-traversing contribution, we use a binned Poisson likelihood to compare our prediction to a baseline expectation.
We compute the number of muons produced by the interactions of muon and tau neutrinos in the vicinity of the detector.
We bin the resulting muons linearly using twenty bins in zenith and logarithmically using two bins per decade in reconstructed energy.
We model the detector energy resolution by introducing an 80\% smearing of the initial energy assuming a normal distribution ~\cite{Aartsen:2013vja}.
We compute the 90\% CL. upper limit on the Ahlers-Halzen GZK model normalization, $\phi_0$, using an Asimov data set, i.e., we set the observed number of events to be the expected mean number of events in the absence of a GZK component.
%Our results are shown for one, five, and ten years of data taking in Tbl.~\ref{tab:table1}.
%Numbers less than one in Table.~\ref{tab:table1} implies that the baseline model can be detected at greater than 90\% CL. and is nominally referred as the model rejection factor.
When considering only Earth-skimming neutrinos, namely declinations such that $\delta < 10^{\circ} $, we get sensitivities that are comparable to current IceCube constraints~\cite{Aartsen:2013vja}.
However, when we include the Earth-traversing component, the sensitivity improves by approximately a factor of two, with most of the improvement coming from the range between 10$^{\circ}$ to 30$^{\circ}$ below the horizon.

\textbf{\textit{Conclusions.\label{sec:conclusions}---}}
The search for UHE neutrinos has been approached using two deeply interconnected techniques: lower-threshold optical neutrino detectors and higher-threshold radio detectors with enhanced sensitivity.
This \textit{letter} demonstrates how the observation of Earth-traversing neutrinos of PeV energy provides information on, and enhanced sensitivity to, their EeV neutrino flux.
We showcase how the technique significantly improves the reach of lower-threshold optical neutrino detectors to transient UHE sources.
We find that the sensitivity to UHE sources is improved by at least an order of magnitude above $10^9$ GeV. Interestingly, optical and radio detectors are more intertwined than originally thought, which implies that the interplay between these energy regimes needs to be taken into account when designing next-generation neutrino telescopes.
In addition, a study of the diffuse flux component demonstrates a gain in sensitivity that is not penalized by the reduced growth of the neutrino cross section with energy~\cite{Arguelles:2015wba}.
However, the observed high-energy astrophysical flux acts as background in this regime, rendering the identification of GZK events possible only by aggregating statistics~\cite{Soto:2021vdc}.
Despite this competition between increasing rate and background, we find that we can improve the current sensitivity by up to a factor of two.

\section*{Acknowledgements}
The authors would like to thank Kohta Murase and John Beacom for useful suggestions and discussions. We also thank Alfonso Garcia Soto and Christopher Wiebusch for providing useful comments on the text.
CAA is supported by the Faculty of Arts and Sciences of Harvard University and the Alfred P. Sloan Foundation.
FH and IS are supported by NSF under grants PLR-1600823 and PHY-1607644 and by the University of Wisconsin Research Council with funds granted by the Wisconsin Alumni Research Foundation.
AK acknowledges the support from the Institute for Gravitation and the Cosmos by the IGC fellowship award.

\bibliography{eev_to_pev}

%%%%%%%%%% supplemental materials %%%%%%%%%%
\pagebreak
\clearpage

%%%%%% SUPLEMENTARY MATERIAL STARTS HERE
%%%%%% SUPLEMENTARY MATERIAL STARTS HERE
%%%%%% SUPLEMENTARY MATERIAL STARTS HERE
%%%%%% SUPLEMENTARY MATERIAL STARTS HERE
\begin{comment}
\onecolumngrid
\appendix

\ifx \standalonesupplemental\undefined
\setcounter{page}{1}
\setcounter{figure}{0}
\setcounter{table}{0}
\setcounter{equation}{0}
\fi

\renewcommand{\thepage}{Supplemental Material-- S\arabic{page}}
\renewcommand{\figurename}{SUPPL. FIG.}
\renewcommand{\tablename}{SUPPL. TABLE}

\renewcommand{\theequation}{A\arabic{equation}}
\clearpage

\begin{center}
\textbf{\large Supplemental Material}
\end{center}
\end{comment}
\end{document}